# INTEGRAL-BALANCE SOLUTION TO THE STOKES' FIRST PROBLEM OF A VISCOELASTIC GENERALIZED SECOND GRADE FLUID

by

*Jordan HRISTOV*

Department of Chemical Engineering

University of Chemical Technology and Metallurgy

1756 Sofia, 8 Kl. Ohridsky Blvd., Bulgaria, e-mail: jordan.hristov@mail.bg

*Integral balance solution employing entire domain approximation and the penetration dept concept to the Stokes' first problem of a viscoelastic generalized second grade fluid has been developed. The solution has been performed by a parabolic profile with an unspecified exponent allowing optimization through minimization of the $L_2$ norm over the domain of the penetration depth. The closed form solution explicitly defines two dimensionless similarity variables $\xi = y/\sqrt{vt}$ and $D_0 = \chi^2 = \sqrt{p/vt^\beta}$, responsible for the viscous and the elastic responses of the fluid to the step jump at the boundary. The solution was developed with three forms of the governing equation through its two dimensional forms (the main solution and example 1) and the dimensionless version showing various sides of the flow field and how the dimensionless groups control it: mainly the effect of the Deborah number. Numerical simulations demonstrating the effect of the various operating parameter and fluid properties on the developed flow filed have been performed.*

**Keywords:** Stokes' first problem, viscoelastic, generalized second grade fluid, Integral-balance solution, Deborah number

**Introduction**

Start-up flows of viscoelastic are intensively modeled in literature under different boundary conditions, flow geometries and constitutive equations [1-6] by analytical and numerical methods. There are very few cases in which their exact analytical solutions can be obtained mainly due to non-linearity imposed by the non-Newtonian constitutive equations of the fluid rheology. The second grade fluid is the common non-Newtonian viscoelastic fluid in industrial fields, such as polymer solution [7], emulsions [8], crude oils [9] extrusion processes [10-12], blood flow [13,14] and magneto-hydrodynamic flows with heat and mass transfer [5].

Fractional calculus allows by replacing the time derivative of an integer order by the fractional order time derivative (in the Riemann-Liouville sense) to incorporate memory effects in the constitutive equations [16]. The second grade generalized fluids with fractional derivatives under start-up flow conditions have been intensively studied toward development of exact analytical solutions [4, 5, 17, 18]. The common approach in those studies is the Laplace transform and then expressions through the generalized Mittag-Leffler function. Even though these solutions are successful they are cumbersome and do not allow easily to detect the contributions of the viscous and elastic effects on the flow development, for example. This article develops an approximate integral balance solution of a fractional generalized second fluid with a special emphasis on simplicity and physical adequacy, without significant loss of exactness. Most of the above mentioned studies are commented further in this work parallel to development of the solution at issue.

**Problem statement**

*Governing equations*

In this article the start-up flow due to a wall suddenly set in motion for a generalized second grade fluid is solved approximately by an integral balance method. For the second grade fluid, the Cauchy stress tensor $\mathbf{T}$ can be expressed as [19, 20]

$$\mathbf{T} = -\rho\mathbf{I} + \mu\mathbf{A}_1 + \alpha_1\mathbf{A}_2 + \alpha_2\mathbf{A}_1^2 \tag{1}$$

where $\rho$ is the density, $\mathbf{I}$ is the unit vector, $\alpha_1$ and $\alpha_2$ are the normal stress moduli, $\mathbf{A}_1$ and $\mathbf{A}_2$ are the kinematical tensors defined as [19]

$$\mathbf{A}_1 = grad\,\mathbf{V} + (grad\,\mathbf{V})^T, \quad \mathbf{A}_2 = \frac{\partial \mathbf{A}_1}{\partial t} + \mathbf{A}_1(grad\,\mathbf{V}) + (grad\,\mathbf{V})^T \mathbf{A}_1 \tag{2a,b}$$

Here, $\partial/\partial t$ denotes a material time derivative, $\mathbf{V}$ is the velocity,

For the second grade fluids with a stress tensor expressed by Eq.(1) that is thermodynamically compatible the following restrictions of the material moduli hold [19]

$$\mu \geq 0, \quad \alpha_1 \geq 0, \quad \alpha_1 + \alpha_2 = 0 \tag{3}$$

Generally, the constitutive relationship of the second grad fluids can be expressed by Eq(1) but the $\mathbf{A}_2$ is defined as

$$\mathbf{A}_2 = D_t^\beta \mathbf{A}_1 + \mathbf{A}_1(grad\,\mathbf{V}) + (grad\,\mathbf{V})^T \mathbf{A}_1, \quad D_t^\beta = \frac{1}{\Gamma(1-\beta)}\frac{d}{dt}\int_0^t \frac{f(\tau)}{(t-\tau)^\beta}d\tau, \quad 0 < \beta < 1 \tag{4a,b}$$

Hence, $D_t^\beta$ denotes a material time derivative of fractional order. With $\beta = 1$ eq. (4a) reduces to Eq.(2b). With $\beta = 0$ and $\alpha_1 = 0$ we get a classical Newtonian liquid.

In absence of body forces, the momentum and continuity equations are

$$\rho\frac{D\mathbf{v}}{Dt} = \nabla \cdot \mathbf{T}, \quad \nabla \cdot \mathbf{V} = 0 \tag{5a,b}$$

*Stokes' first problem of a second grade fractional (viscoelastic) fluid*

Consider a semi-infinite space filled by a generalized second grade viscoelastic fluid undergoing a transient motion due top placed infinite plate suddenly set in motion with a velocity $U_0$ (parallel to the $x$ axis). That is, the velocity field is $\mathbf{U} = u(y,t)$, where $y$ is the axis normal to plate surface and $u$ is the velocity component along the $x$ axis. In this context, the stress tensor component relevant to problem at issue is

$$\mathbf{T}_{xy} = \mu \frac{\partial u}{\partial y} + \alpha_1 D_t^\beta \frac{\partial u}{\partial y} \tag{6}$$

Commonly, the constitutive equation for second grade liquid is expressed in two forms [21]

$$\tau(t) = \mu \varepsilon(t) + E \frac{d\varepsilon(t)}{dt} \tag{7a}$$

$$\tau(t) = \mu \varepsilon(t) + \alpha_1 D_t^\beta \left[ \varepsilon(t) \right] \tag{7b}$$

Moreover, $T_{xx} = T_{yy} = T_{zz} = T_{xz} = T_{yz} = 0$, where $T_{xy} = T_{yx}$.

Therefore, in (6) the coefficient $\alpha_1$ is the first normal stress modulus, while the Riemann-Liouville operator $D_t^\beta$ denotes this as a model of a general second grade fluid. The operator $D_t^\beta$ has a fractional time dimension $\left[ \rho y^2 / \mu \right]^\beta$. Thus, $\alpha_1$ in Eq.6 and Eq.7b equals the viscoelastic coefficient $E$ (see Eq.7a) with a time divisor $c$ (matching constant), $\alpha_1 = E/c$, where $c = \left[ \rho y^2 / \mu \right]^{\beta-1}$.

Therefore, the equation of motion becomes

$$\rho \frac{\partial u}{\partial t} = \mu \frac{\partial^2 u}{\partial y^2} + \alpha_1 D_t^\beta \frac{\partial^2 u}{\partial y^2} \tag{8}$$

with the following boundary and initial conditions

$$u(y,0) = 0, \ y > 0; \quad u(0,t) = U_0, \ t > 0; \quad u \to 0, \ y \to \infty \tag{9a,b,c}$$

**Integral balance approach to the Stokes' first problem**

The approach consider a velocity $u(y,t)$ over a finite penetration depth $\delta(t)$ [22-25] (see also [26] for illustrative examples of this concept). At the font of the penetration depth we have

$$u_{y=\delta} = 0 \text{ and } \left. \frac{\partial u}{\partial y} \right|_{y=\delta} = 0. \tag{10a,b}$$

Therefore, the condition (10) replaces (9c) in the set of boundary conditions, considering an undisturbed fluid at $y \geq \delta$.

The profile, can be expressed as an entire domain function

$$u = b_0 + b_1 \left( 1 - \frac{y}{\delta} \right)^n \tag{11}$$

which is a general parabolic profile with unspecified exponent $n$.

Applying the boundary conditions (9a,b) and (10b) to the profile (11) we get

$$y = 0, \; u = U_0 = b_0 + b_1 \tag{11a}$$

$$y = \delta, \; u = 0 = b_0 \tag{11b}$$

$$y = \delta, \; \left.\frac{\partial u}{\partial y}\right|_{y=\delta} = \frac{n}{\delta}\left(1-\frac{y}{\delta}\right)^{n-1} = 0 \tag{11c}$$

The condition (11c) is valid for any value of the exponent $n$ [24,25,27]. Therefore, the approximate profile is

$$u = U_0\left(1-\frac{y}{\delta}\right)^n \Rightarrow \frac{u}{U_0} = \left(1-\frac{y}{\delta}\right)^n \tag{12a}$$

Equation (8) can be expressed as

$$\frac{\partial u}{\partial t} = \nu \frac{\partial^2 u}{\partial y^2} + pD_t^\beta \frac{\partial^2 u}{\partial y^2} \tag{13}$$

where $p = \alpha_1/\rho \left[m^2\right]$. It is worth nothing, but in some articles $p$ is expressed as $p = \nu\lambda_r$, where $\lambda_r$ is the relaxation time.

Integrating (13) from $0$ to $\delta$ we get

$$\int_0^\delta \frac{\partial u}{\partial t}dy = \int_0^\delta \nu\frac{\partial^2 u}{\partial y^2}dy + \int_0^\delta pD_t^\beta \frac{\partial^2 u}{\partial y^2}dy \tag{15a}$$

The LHS of (15a,b) is the *momentum integral balance* which can be expressed by the Leibnitz rule, namely

$$\int_0^\delta \frac{\partial u}{\partial t}dy = \frac{d}{dt}\int_0^\delta u\,dy - u|_{y=\delta}\frac{d\delta}{dt} \tag{15b}$$

**Penetration depth**

*Integration without scaling of the governing equation*

The integration in (15a), applying the Leibnitz rule to the LHS (15b) and with $u = (1-y/\delta)^n$ under the integrals of all terms in (15a) (see [24,27-30] about this technique), we get an equation about the time evolution of the penetration depth $\delta$, namely

$$\frac{d\delta^2}{dt} = 2\nu n(n+1) + 2p\frac{n(n+1)}{(1-\beta)\Gamma(1-\beta)}t^{-\beta} \tag{16}$$

The solution of (16), with the initial condition $\delta(0) = 0$ is

$$\delta^2 = 2\nu n(n+1)t + 2\frac{pn(n+1)}{(1-\beta)\Gamma(1-\beta)}t^{1-\beta} \Rightarrow \delta = \sqrt{2\nu n(n+1)t + 2pn(n+1)t\left(\frac{j_\beta}{t^\beta}\right)} \tag{17a}$$

$$j_\beta = \frac{1}{(1-\beta)\Gamma(1-\beta)} \tag{17b}$$

Re-arranging the second form about $\delta$ in (17a) we get

$$\delta = \sqrt{\nu t}\sqrt{2n(n+1)\left[1+\frac{p}{\nu t^{\beta}}j_{\beta}\right]} \qquad (18)$$

Obviously, from Eq.18 with $p=0$ or $p/\nu = \alpha_1/\mu \to 0$, or more realistic as the time goes on, we get the Newtonian penetration depth (integral balance solution), i.e. $\delta = \sqrt{\nu t}\sqrt{2n(n+1)}$ [24,25]. Hence, the momentum impulse penetrates into the fluid at a depth $\delta \equiv \sqrt{\nu t}$, a fact well-known from the Rayleigh solution of the Stokes's first problem [26].

Denoting $p/\nu = t_{ev}$ we have $t_{ev}/t^{\beta} = D_0$ (see further in this work comments about the definition of this ratio), we get

$$\delta = \sqrt{\nu t}\sqrt{2n(n+1)\left[1+j_{\beta}D_0\right]} \qquad (19)$$

It is evident, that at a given fractional order $\beta$, the contribution of the elastic properties of the fluid to the penetration depth depends on the magnitude of the product $j_{\beta}D_0$. Hence, the limit of impact of the elastic term on the approximate profile can be assumed at $j_{\beta}D_0 \approx 10^{-3}$, for example.

*Integration after scaling of the governing equation*

Now, let us consider how the rescaling and the dimensionless form of Eq. (8) affect the outcome of the integral balance approach. Two examples taken from the literature will be used for this purpose next.

- **Example 1:** Scaling (8) by $\bar{y} = y/\sqrt{p}$ and $\bar{u} = u/U_0$ [6] we get

$$\frac{\partial \bar{u}}{\partial t} = \left(\frac{\nu}{p}\right)\frac{\partial^2 \bar{u}}{\partial \bar{y}^2} + D_t^{\beta}\frac{\partial^2 \bar{u}}{\partial \bar{y}^2} \Rightarrow \frac{\partial \bar{u}}{\partial t} = \left(\frac{\nu}{p}\right)\frac{\partial^2 \bar{u}}{\partial \bar{y}^2} + D_t^{\beta}\frac{\partial^2 \bar{u}}{\partial \bar{y}^2} \qquad (20)$$

where the ratio $p/\nu = \left[s^{-\beta}\right]$ is inverse of the fractional relaxation time.
Then, using the same approximation profile $\bar{u} = u/U_0 = \left(1-\bar{y}/\bar{\delta}\right)^n$ where $\bar{\delta} = \delta/\sqrt{p}$ and performing an integration of (20) from 0 to $\delta$ we get

$$\frac{d}{dt}\left(\frac{\bar{\delta}}{n+1}\right) = \left(\frac{\nu}{p}\right)\frac{n}{\bar{\delta}} + \frac{n(n+1)}{(1-\beta)\Gamma(1-\beta)}\frac{d}{dt}\left(t^{1-\beta}\right) \qquad (21a)$$

Then, the equation governing the evolution of the penetration depth is

$$\frac{d\bar{\delta}^2}{dt} = 2\left(\frac{\nu}{p}\right)n(n+1) + 2j_{\beta}n(n+1)\frac{d}{dt}\left(t^{1-\beta}\right) \qquad (21b)$$

Taking into account the initial condition $\delta(t=0)=0$, from the integration of (21b) we get

$$\bar{\delta} = \sqrt{\frac{v}{p}} t \sqrt{2n(n+1)\left[1 + j_\beta \frac{p}{v t^\beta}\right]} \tag{22}$$

The second expression in (22) is equivalent to (18) and (19) taking account the rescaling of the space co-ordinate by $1/\sqrt{p}$.

- **Example -2:** Making eq. (18) dimensionless [3] by $u^* = u/U_0$, $y^* = yU_0/v$, $t^* = tU_0^2/v$, $\eta = \frac{\alpha_1 U_0^2 \rho}{\mu^2}$ we get

$$\frac{\partial u^*}{\partial t^*} = \frac{\partial^2 u^*}{\partial y^{*2}} + \eta D_t^\beta \frac{\partial^2 u^*}{\partial y^{*2}} \tag{23}$$

$$\frac{d\delta^{*2}}{dt} = 2n(n+1) + 2\eta n(n+1) j_\beta \frac{d}{dt^*}\left(t^*\right)^{1-\beta} \tag{24}$$

Then the dimensionless penetration depth $\delta^*$ is

$$\delta^* = \sqrt{2n(n+1)t^*\left[1 + \eta \frac{j_\beta}{\left(t^*\right)^\beta}\right]} \;,\; \eta = \frac{\alpha_1 U_0^2 \rho}{\mu^2} = \frac{(p/v)}{(v/U_0^2)} = \frac{t_{pv}\,(\text{relaxation})}{t_v\,(\text{newtonian momentum diffusion})} \tag{25a,b}$$

*Some comments on the penetration depth and the dimensionless numbers*

The expressions of the penetration depth developed contain terms related to the elastic component of the stress tensor, which logically decay in time. In those terms the control on the magnitude of the elastic contribution to the momentum penetration depth is through $j_\beta D_0$ or $\left(t^*\right)^\beta = \left(tU_0^2/v\right)^\beta$. In the former case *the intrinsic time scale* is defined by the ratio $p/v$ and to some intent $D_0 = p/vt^\beta$ addresses the general definition of the Deborah number. Moreover, in Example 2, the time scale is defined by the boundary condition at $x = 0$ and the diffusion coefficient of the Newtonian momentum flux $v$ defined by first term of the constitutive relationship (7b), i.e. $v/U_0^2\,[s]$. Hence, $\eta \equiv D_0$ because from (25b), we have,

$$\eta = \left(p/vt^\beta\right)\left(U_0^2 t^\beta/v\right) = D_0\left(U_0^2 t/v\right)t^{\beta-1} = D_0\, t^* t^{\beta-1}. \tag{26}$$

Recall, *the first Stokes' problem in a semi-infinite fluid has no a characteristic length scale*, in contrast to the Couette flow. To some extent, using $l = \sqrt{vt}$ as a length scale (see the classical solutions of transient heat conduction problems [31] in the Newtonian flow), we get mechanistically $l_\beta = \sqrt{vt^\beta}$ a fractional length scale [29, 30], that is the ratio $p/vt^\beta$ is dimensionless. In this context, the penetration depth ca be expressed in two equivalent forms, namely

$$\delta^* = \sqrt{2n(n+1)t^*\left[1 + j_\beta D_0 \left(t^*\right)^{1-\beta}\right]} \Rightarrow \delta^* = \sqrt{\eta t^*}\sqrt{\frac{2n(n+1)}{\eta}\left[1 + \frac{j_\beta}{\left(t^*\right)^\beta}\right]} \tag{27a,b}$$

The expression (27a,b) is equivalent to (19) and (22) taking into account $\delta = \delta^* (\nu/U_0)$ and the dimensionless time defined by $t^* = tU_0^2/\nu$.

Taking into account that the time ratio $(p/\nu)/t$ emerges in the expression of the penetration depth, we refer to the *Elasticity number*, $El$ defiend as $El = \alpha_1/(L^2/\nu)$, where $L^2/\nu$ is the convective (Newtonian) macroscopic time scale. Assuming $L = l_{\beta_2} = \sqrt{\nu t^\beta}$ in the case of viscoelastic flow we get $El = \alpha_1/t^\beta$. In this context, $D_0 = p/\nu t^\beta = \left(\sqrt{p}/\sqrt{\nu t^\beta}\right)^2 = L_p/l_\beta$, where $L_p = \sqrt{p}$. Taking into account the comments about eq. (13) we can express $p$ as $p = \nu\lambda_r$ and get $L_p = \sqrt{\nu\lambda_r}$. The length scale ratio $D_0 = L_p/l_\beta$ can be considered as a proportion of the depths of penetration of the impulse imposed at the boundary and transported into the fluid by two mechanisms: *the viscous shear* $(l_\beta)$ and *the elastic response* $(L_p)$. Small $D_0$ by reason of low relaxation time $\lambda_r$ or as the time goes on, shifts the fluid flow towards Newtonian one. Otherwise, with increase in $p$ caused by large relaxation times $\lambda_r$ or due to short $l_\beta$ (high viscosity $\mu$) the fluid behaves as a solid. In any case, all these modes depend on the time scale chosen, which is the crux in the definition of the Deborah number.

In addition to above comments on the physical meaning of $D_0 = p/\nu t^\beta = \left(\sqrt{p}/\sqrt{\nu t^\beta}\right)^2$, it can be considered as a similarity variable with a length scale $\sqrt{\nu t^\beta}$ and a fixed space co-ordinate $\sqrt{p}$. Hence, the approximate velocity profile is a function of two similarity variables: $\xi = y/\sqrt{\nu t}$ and $\chi = \sqrt{p}/\sqrt{\nu t^\beta}$. The former corresponds to the Newtonian flow (see the next section) while the second similarity variable corresponds to the elastic mode of momentum transport ($\chi^2 = D_0$), depends on the fractional order $\beta$ and decays in time.

Because the approximate integral-balance solution of the Newtonian flow is straightforward, it is possible to extract the term representing only shear momentum penetration depth $\delta_N = \sqrt{\nu t}\sqrt{2n(n+1)}$. This allows writing the penetration depth in a dimensionless form $\delta/\delta_N$, namely:

$$\Delta = \frac{\delta}{\sqrt{\nu t}\sqrt{2n(n+1)}} = \sqrt{1 + j_\beta D_0} \qquad (28a)$$

Similarly, for examples 1 and 2 we get, $\Delta_1$ and $\Delta_2$, namely

$$\Delta_1 = \frac{\overline{\delta}}{\sqrt{D_0 t}\sqrt{2n(n+1)}} = \sqrt{[1 + j_\beta D_0]}, \quad D_0 = \nu/p \left[s^{-1}\right] \qquad (28b)$$

$$\Delta_2 = \frac{\delta^*}{\sqrt{\eta t}\sqrt{\frac{2n(n+1)}{\eta}t^*}} = \sqrt{1 + \frac{j_\beta}{(t^*)^\beta}} \qquad (28c)$$

It is evident that, we have only viscous (Newtonian) flow when the $\Delta$, $\Delta_1$ and $\Delta_2$ become equal to unity, i.e. $D_0 \to 0$. **G**raphical examples in Fig. 1a, b, c show some numerical experiments illustrating the effects of the elastic terms on the time evolution of the relative penetration depth $\Delta_2$. Taking into account the elastic contribution to the momentum penetration depth decays in time, all plots exhibits high values at small times, but we have to bear in mind that the magnitude of the ratio $p/\nu$, i.e., the retardation time determine how much relative penetration depth would exceed unity, i.e. the scale defined by the purely viscous flow.

**Velocity field**

Therefore, with the expression of $\Delta$, for example, the velocity profile is,

$$\frac{u}{U_0} = \left(1 - \frac{y}{\sqrt{vt}} \frac{1}{\sqrt{2n(n+1)(1+j_\beta D_0)}}\right)^n = \left(1 - \xi \frac{1}{\sqrt{2n(n+1)(1+j_\beta \chi^2)}}\right)^n = \left(1 - \frac{\xi}{F_n R_\beta}\right)^n \quad (29a,b)$$

where $\xi = y/\sqrt{vt}$ is the Boltzmann similarity variable with $\sqrt{vt}$ as a diffusion length scale (see the comments above)., $F_n = \sqrt{2n(n+1)}$ and $R_\beta = 1 + j_\beta D_0$ ($R_\beta = 1 + j_\beta \chi^2$). In the Newtonian problem the exact solution is $u/U_0 = 1 - erf(\xi/2)$ with $0 \leq \xi \leq \infty$. In the integral-balance solution of the Newtonian problem we have $0 \leq \xi \leq F_n$, while for the viscoelastic flow the range is $0 \leq \xi \leq F_n R_\beta$. The increase in the fluid retardation time, results in increased length scale of the integral balance penetration depth by a factor of $R_\beta$ with respect to that defiend by the shear flow only.

**Calibration of the exponent**

*Newtonian flow*

First we can consider the limiting case $D_0 = 0$, which has been analyzed thoroughly in its heat-diffusion version [27,28, 32, 33]. With calibration at $y = 0$ [24], we have $n = 1.675$. However, the minimization of the $L_2$ norm in the domain $0 \leq y \leq \delta$ results in $n \approx 2.35$ applying the method of Myers [33] and $n \approx 1.507$ applying a similarity transform of eq. (13) at $p = 0$ [32]. Moreover, the self-adaptive exponent concept [34] allows the initial exponent value $n_0$ to be defined through a fractional (half-time) condition at $y = 0$. Precisely, this concept uses $n = n_0 + k_j LambertW(\xi)$, with $n_0 = 1.675$ corresponding to the problem with problem with $p = 0$ [24].

*From Newtonian to Viscoelastic flow*

The general approach in estimation of the optimal exponent is to minimize the $L_2$ norm in the domain $0 \leq y \leq \delta$, namely

$$E(y, t, \beta) = \int_0^\delta \left[\frac{\partial \bar{u}}{\partial t} - v \frac{\partial^2 \bar{u}}{\partial y^2} - p D_t^\beta \frac{\partial^2 \bar{u}}{\partial y^2}\right]^2 dy \to \min \quad (30)$$

With $\bar{u} = (1 - y/\delta)^n$ we have

$$\frac{\partial \bar{u}}{\partial t} = n\left(1 - \frac{y}{\delta}\right)^{n-1}\left(-\frac{y}{\delta^2}\right)\frac{d\delta}{dt}, \quad \frac{\partial^2 \bar{u}}{\partial y^2} = \frac{n(n-1)}{\delta^2}\left(1 - \frac{y}{\delta}\right)^{n-2} \quad (31a,b)$$

$$D_t^\beta \frac{\partial^2 \bar{u}}{\partial y^2} = \frac{1}{(1-\beta)\Gamma(1-\beta)} \frac{n(n-1)}{\delta^2}\left(1 - \frac{y}{\delta}\right)^{n-2} \frac{1}{t^\beta} \quad (32)$$

$$\frac{d\delta}{dt} = -\sqrt{2n(n+1)} \frac{1 - (1-\beta)D_0}{(t + j_\beta D_0)^{3/2}}, \quad \delta \frac{d\delta}{dt} = \sqrt{vt}\left[2n(n+1)\right]\left[1 + j_\beta D_0\right]\left[1 - (1-\beta)D_0\right], \quad (33a,b)$$

After integration in (30), $E(y,t,\beta)$, following the idea of Langford [35](see also [27-30] and [33] ) can be expressed as $E(y,t,\beta) = \frac{1}{(vt)^{3/2}} e_n(y,t,\beta) \equiv \frac{1}{\delta^3} e_n(y,t,\beta)$. Taking into account that the entire function $E(y,t,\beta)$ decays in time with a speed $\equiv \delta^3$, the minimization refers to an optimal value of $n$ minimizing the function.

$$e_n(y,t,\beta) = \delta^4 \frac{d\delta}{dt} \frac{n}{(2n-1)(2n+1)} + v^2 \frac{(n-1)}{2} + v\delta \frac{d\delta}{dt} \frac{n^2(n-1)}{2n-1} + \\ + \delta^3 \frac{j_\beta}{t^\beta} \frac{n^2(n-1)}{2n-1} - 2p \frac{j_\beta}{t^\beta} \delta \frac{d\delta}{dt} \frac{n(n-1)}{2} - \frac{vp}{t^\beta} n^2(n-1) \tag{34}$$

According to Myers [33], setting all time-dependent terms (increasing in time only) of (34) equal to zero, we get

$$e_n(y,\beta,t=0) = v^2 \frac{(n-1)}{2} - \frac{vp}{t^\beta} n^2(n-1) \;, \quad n = \frac{1}{\sqrt{2}} \frac{1}{\sqrt{D_0}} \tag{35a,b}$$

A simple test with $n=2$ and $n=3$ provides $D_0 \approx 0.125$ and $D_0 \approx 0.055$, respectively. In this context, the optimal Newtonian profiles have exponents: $n \approx 2.35$ corresponds to $D_0 \approx 0.100$, while that $n \approx 1.507$ provides $D_0 \approx 0.220$, i.e. flows with weak viscoelastic effects (low Deborah numbers). Therefore, within the range defined by $O(D_0) \sim 1$, that is the viscoelastic flow [37], the exponent of the parabolic profile oscillates around $n=2$, as the value of $D_0$ varies. Taking into account, the inverse time dependence of $D_0$, it is reasonably to expect an increase in the value of the exponent $n \sim t^{\beta/2}$ as it follows from (35b). That is, the shorter relation times (i.e. larger observation times), the larger exponents of the approximate profile and *vice versa*. However, if we do not look for an optimal value of $n$ but apply the classical heat-balance integral approach [23, 27], with either $n=2$ or $n=3$ we get

$$\frac{u}{U_0} \approx \left(1 - \frac{y}{3.464\sqrt{vt}\sqrt{1+j_\beta D_0}}\right)^2 = \left(1 - \frac{\xi}{3.464\sqrt{1+j_\beta \chi^2}}\right)^2 \tag{36a}$$

$$\frac{u}{U_0} \approx \left(1 - \frac{y}{4.898\sqrt{vt}\sqrt{1+j_\beta D_0}}\right)^3 \approx \left(1 - \frac{\xi}{4.898\sqrt{1+j_\beta \chi^2}}\right)^3 \tag{36b}$$

Moreover, applying the profile with a self-adaptive exponent [34] we get

$$\frac{u}{U_0} \approx \left(1 - \frac{\xi}{\left[n_0 + k_j LambertW(\xi)\right]\sqrt{1+j_\beta \chi^2}}\right)^{\left[n_0 + k_j LambertW(\xi)\right]} \;, \quad n_0 = 1.675, \; k_j = 0.5 \tag{36c}$$

In fact, the concept of the self-adaptive exponent implies: the shorter observation times, the lower exponents and *vice versa*. This corresponds to the general tendency $n \sim D_0^{-1/2} \equiv t^{\beta/2}$ drawn above. The example shown in Fig.1d really indicates that discrepancies in the approximate profiles become significant beyond $\xi \approx 1.5$. More precisely, in the range $0 \leq \xi \leq 1$ the profiles with all exponents mentioned above are practically indistinguishable.

**Stress field**

The stress field can be represented in accordance with the versions discussed above

$$F = \nu \frac{\partial u}{\partial y} + p D_t^\beta \left(\frac{\partial u}{\partial y}\right) \Rightarrow F(y,t) = \nu \frac{n}{\delta}\left(1-\frac{y}{\delta}\right)^{n-1}\left[1 + j_\beta \frac{p}{\nu t^\beta}\right], \quad F(t, y=0) = \nu \frac{n}{\delta}\left[1 + j_\beta \frac{p}{\nu t^\beta}\right] \quad (37a,b,c)$$

*Example 1:*

$$F_1 = \left(\frac{\nu}{p}\right)\frac{\partial \bar{u}}{\partial \bar{y}} + D_t^\beta \left(\frac{\partial \bar{u}}{\partial \bar{y}}\right) \Rightarrow F_1(y,t) = \frac{n}{\delta}\left(1-\frac{y}{\delta}\right)^{n-1}\left[\frac{\nu}{p} + j_\beta \frac{1}{t^\beta}\right], \quad F_1(t, y=0) = \frac{n}{\delta}\left[\frac{\nu}{p} + \frac{j_\beta}{t^\beta}\right] \quad (38a,b,c)$$

*Example 2:*

$$F_2 = \frac{\partial u}{\partial y} + \eta D_t^\beta \left(\frac{\partial u}{\partial y}\right) \Rightarrow F_2(y,t) = \frac{n}{\delta}\left(1-\frac{y}{\delta}\right)^{n-1}\left[1 + \eta \frac{j_\beta}{t^\beta}\right], \quad F_2(t, y=0) = \frac{n}{\delta}\left[1 + \eta \frac{j_\beta}{t^\beta}\right] \quad (39a,b,c)$$

where $F_2 = \mathbf{T}_{xy}/\rho U_0^2$

The expressions (37)-(39) clearly indicate that the stress at a given point, at any time, is dependent on the time history of the velocity profile at that point; the time history is presented by the fractional derivative term associated with elastic response of the fluid. Numerical experiments are commented in the next section.

**Numerical experiments**

The plots in Fig. 2 reveal that with increases in $D_0 = \chi^2$ the profile $u/U_0 = f(\xi, \chi = const.)$ becomes almost linear as $D_0$ increases. Hence, with increase in $D_0$, the exponent $n$ should decreases from the initial values established for the pure viscous regime. Therefore, to same extent, the relationship (32, b) shows the real tendency even though the values calculated are not exact. The result shown in Fig.2 indicate that with $0 \leq \chi \leq 0.5$ there is no strong effect of the elasticity on the profile.

The short time ($0 \leq t \leq 0.1$) profiles at different point from the moving plate in the depth of the fluid are shown in Fig. 3 as good examples of the effect of magnitude of the ratio $p/\nu$, i.e. the fluid elasticity. Similar effects are shown by the profiles in Fig. 4a (short distance from the plate) and Fig. 4b (large distance from the plate), but at low $p/\nu$ ratios. The variation in both $y$ and $p/\nu$ at short and medium times are well illustrated by the 3D numerical results in Fig.4c,d.

The short-time 3D numerical simulations (see Fig. 5) with variations in the fractional order $\beta$ clearly indicate the effect of the $p/\nu$ ratio. The range $0 \leq D_0 \leq 1$ (Fig. 5a) results in almost parabolic profiles, while increase in $p/\nu$ results in $0 \leq D_0 \leq 5$ (Fig. 5b), $0 \leq D_0 \leq 10$ (Fig.5c) and $0 \leq D_0 \leq 20$ (Fig. 5d) and more or less linear velocity profiles when $\beta \to 1$ (the simulations are performed with $\beta_{max} = 0.9$).

Decrease in the fractional order from $\beta \approx 0.5$ to $\beta \approx 0.1$ yields almost parabolic profiles. The effect of the low fractional order $\beta$ is similar to that observed at low $p/\nu$ ratios and $0 \le D_0 \le 1$ but at $0.5 \le \beta \le 0.9$. These simulations break slightly the well established models and the necessary condition the generalized Oldroyd-B fluid to be thermodynamically compatible and have a mechanical analogue made up of fractional elements: the two orders of fractional derivatives in the constitutive equations are equal [37]. However, the numerical simulations answer to some curiosity beyond the established rules and show what happens if the fractional order is varied.

When the governing equation is transformed into a dimensionless form, as it was demonstrated by Example 2, the effect of the parameter $\eta$, that is equivalent to $D_0$, is the same as that demonstrated above (see Fig. 6). However, in this version of the integral-balance solution the effect of the Deborah number (i.e. $D_0$) is well presented by variations in the velocity profile shapes varying from parabolic (low $D_0$) toward practically linear as $D_0$ increases.

**Discussion**

The integral balance solution provides a simple closed form solution of the first Stokes's problem with a generalized second grade fluid commonly solved by exact methods with cumbersome expressions.

Beyond the correctness and the physical adequacy of both the constitutive relationships and the model, the he main question arising in modelling is: how the final solutions can be used for calculations and analyzed for relationships between the controlling parameters? The integral-balance solution developed explicitly answers this question by clearly defining two dimensionless similarity variables $\xi = y/\sqrt{\nu t}$ and $D_0 = \chi^2 = \sqrt{p/\nu t^\beta}$, responsible for the viscous and the elastic responses of the fluid to the step jump at the boundary.

Moreover, the integral balance solution directly shows:

i) How the term containing $D_0$ decay is time thus shifting the solution towards the Newtonian problem controlled by the fluid kinematic viscosity only?

ii) How the weight of the term containing $D_0$ depends on the fractional order $\beta$?

iii) What is the suitable order of the approximate profile exponent and how it corresponds to the value of $D_0$? In this context, the classical $n = 2$ and $n = 3$, used in the simulation, are adequate because they correspond to two principle conditions: i) to minimize the $L_2$ of the approximate solution and ii) to match the fluid-like behaviour of the viscoelastic medium with low Deborah numbers.

The technology of the integral balance solution to fractional order diffusion equations (diffusion of mass, heat or momentum) is straightforward. It may be performed also by other approximate profile such as polynomial of 2$^{nd}$ or 3$^{rd}$, like its classical version [22, 23]. However, the use of a profile with an unspecified exponent give us a liberty to formulate the problem generally and then to look for the optimal solution. The present solution used a fixed exponent but the problem of variable exponent depending on the Deborah number (a tentative relationship emerging in this study) is still open and has to be resolved.

**Conclusions**

The Stokes' first problem of generalized viscoelastic generalized second grade fluid was solved by approximate integral balance solution employing entire domain parabolic profile with unspecified exponent and the penetration depth concept. The main outcomes of this solution can be outlined as:

1) Simple concept providing a closed form solution with well distinguished terms corresponds to the viscous and elastic effect accordingly.
2) The physical clarity of the developed solution allows to analyze the model and establish the domain of its application (the range defined by the transport coefficients of the constitutive equations) and to demonstrate when it approaches or goes far a way from the classical Newtonian solution, as well.
3) The approach developed in earlier studies employing the same approximate profile for searching the optimal exponent via minimization of the $L_2$ provides a physically adequate result relating the exponent and the Deborah number.
4) **For seek of simplicity of the numerical simulations were performed $n = 2$ and $n = 3$ because the main goal was to show the effect of the competition between the viscous and the elastic effects on the flow field.**
5) **The Integral balance solution was developed with three forms of the governing equation through its two dimensional forms (the main solution and example 1) and the dimensionless version (example 2). These three solutions permit the approximated profile to be expressed through dimensionless parameters and, to some extent, to show various sides of the flow field and how the dimensionless groups control it: mainly the effect of the Deborah number.**

Nomenclature

$b_0$ - coefficient in Eq. 11 , (dimensionless)
$b_1$ - coefficient in Eq. 11 , (dimensionless)
$D_0 = p/\nu t^\beta$ , dimensionless number
$E$ -viscoelastic constant, [ $N.m^{-2}.s^2$ ]
$El = \alpha_1/t^\beta$ -Elasticity number, (dimensionless)
$F$ -stress, [ $N.m^{-2}$ ]
$k_j$ -weight coefficient of the variable exponent (see Eq. 36c), (dimensionless)
$L$ -length scale , [ $m$ ]
$n$ -exponent of the parabolic profile, (dimensionless)
$t$ - time, [ $s$ ]
$y$ -space co-ordinate, [ $m$ ]
$u$ -velocity, [ $m.s^{-1}$ ]

*Greek letters*

$\alpha_1$ -viscoelastic coefficients, [ $N.m^{-2}s^2$ ], $\alpha_1 = E/c$ , where $c = \left[\rho y^2/\mu\right]^{\beta-1}$.
$\beta$ -fractional order (dimensionless)

$\varepsilon(t)$ - strain, [ $s^{-1}$ ]

$\lambda_r$ - relaxation time, [ $s$ ]

$\mu$ - dynamic viscosity [ $N.m^{-2}.s$ ]

$\nu$ - kinematic viscosity [ $m^2.s^{-1}$ ]

$\rho$ - density, [ $kg.m^{-3}$ ]

$\tau$ - shear stress, [ $N.m^{-2}$ ]

Subscripts

1- Example 1

2- Example 2

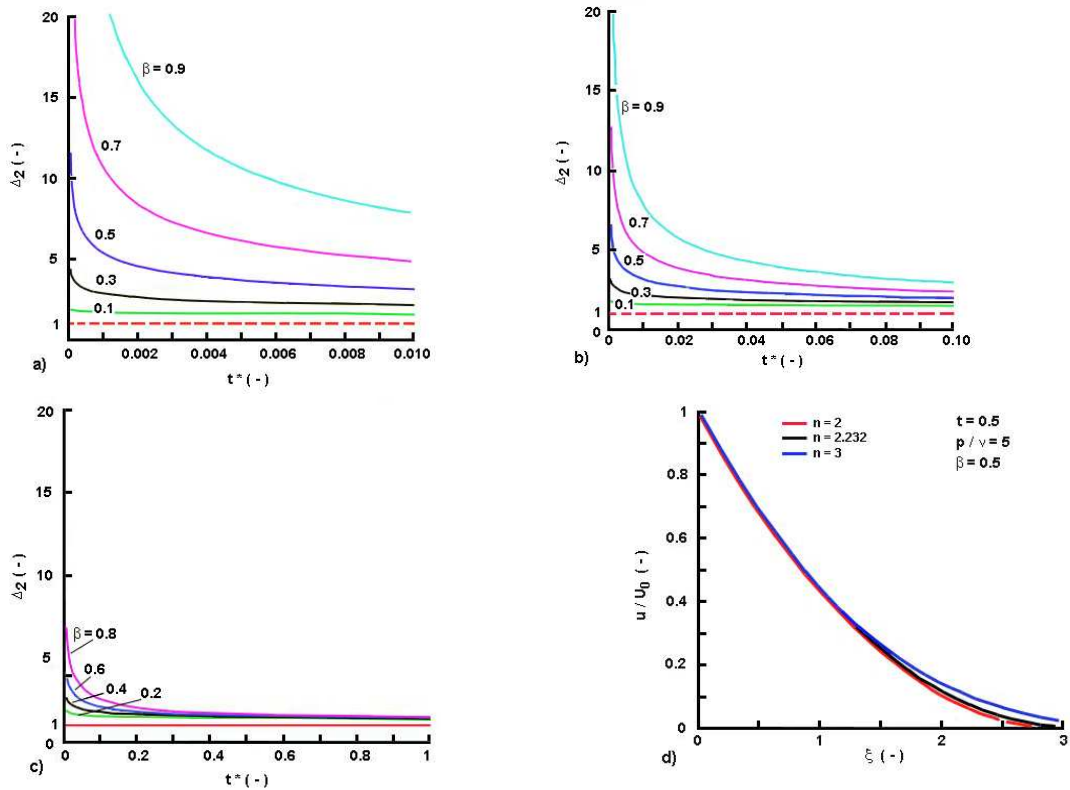

Fig. 1. Relative penetration depth and approximate velocity profiles (the case of Example 2).
a,bc) Relative penetration depth as function of the dimensionless times $t^* = tU_0^2/\nu$ : effect of the and the fractional order $\beta$ :a) Short dimensionless times , b) Medium dimensionless times , c) Large dimensionless times
d) Dimensionless velocity profile with different exponents ; $\xi = y/\sqrt{\nu t}$ .

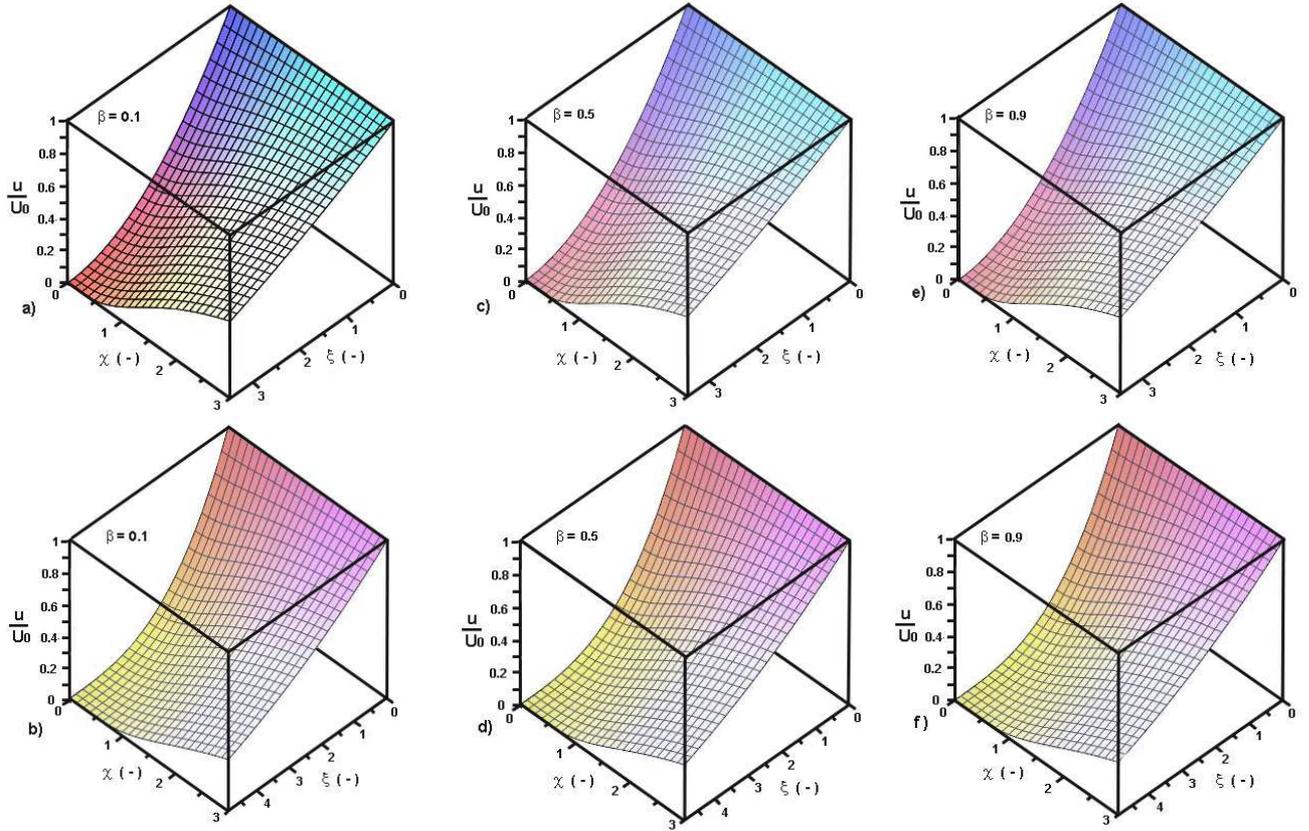

Fig. 2. Velocity field as a function of two similarity variables: $\xi = y/\sqrt{\nu t}$ and $\chi = \sqrt{p/\nu t^\beta}$, and the fractional order $\beta$ as a parameter. The velocity profile is expressed by Eq. (29) with various exponents.
Left column (a, b): profiles with $\beta = 0.1$;
Middle column (c, d): profiles with $\beta = 0.5$;
Right column (e, f): profiles with $\beta = 0.9$;

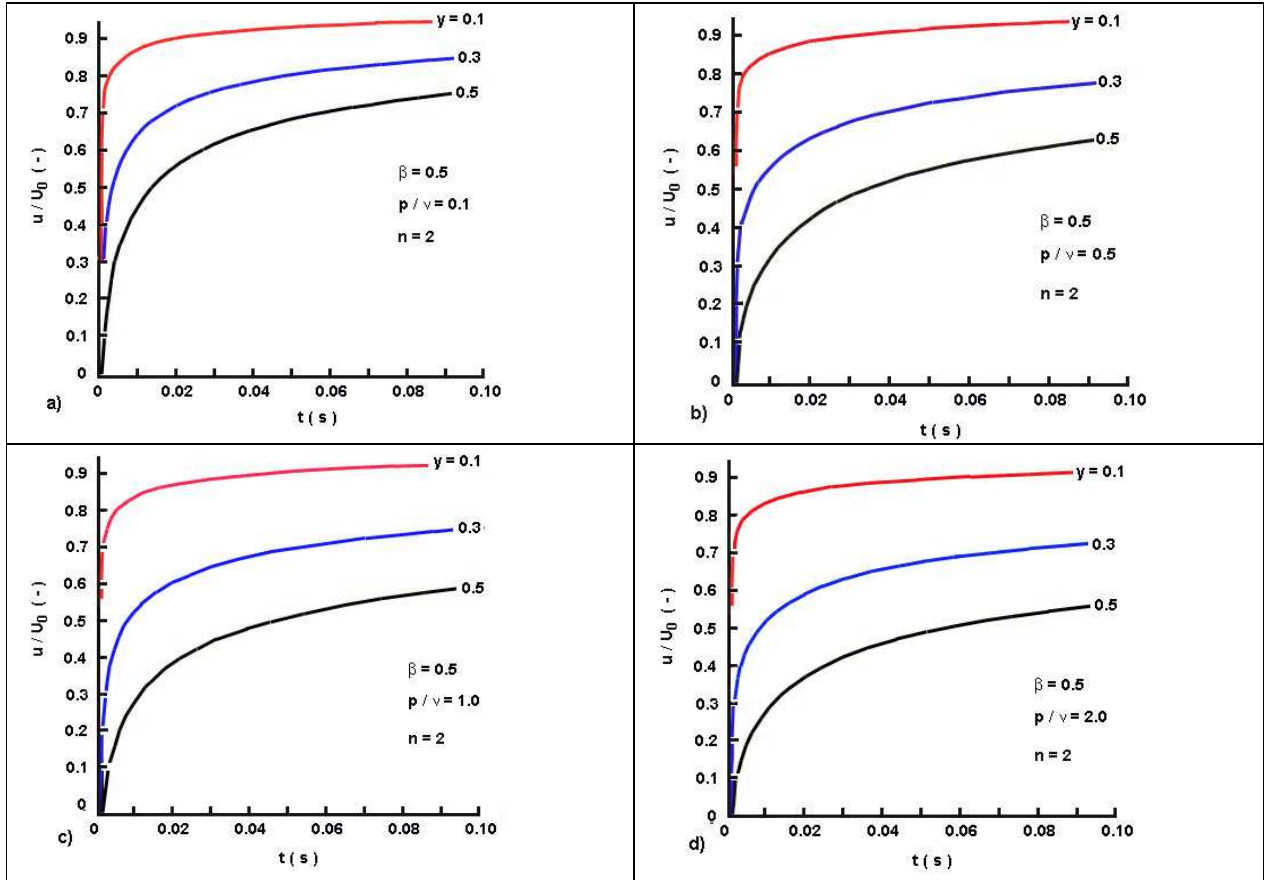

Fig. 3. Velocity as a function of the time (short time range) at fixed distances from the plate ( $y = 0$ ) represented by an approximate profile with $n = 2$ and $\beta = 0.5$. These plots show the effect of the increased ratio $p/\nu$ on the velocity field. Example 1.

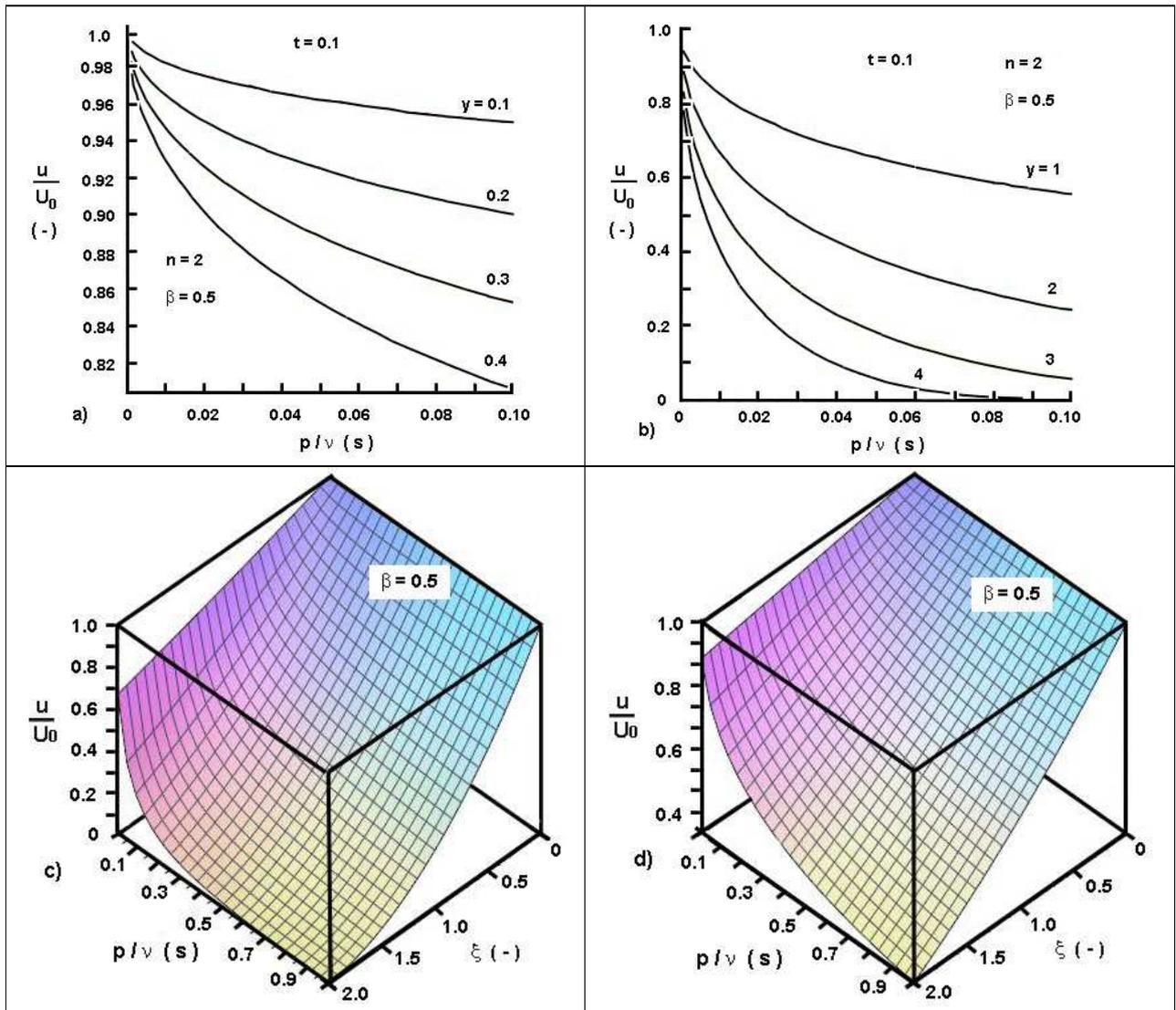

Fig. 4. Velocity field is represented by the penetration depth expressed by eq.(22) : $n = 2$ and $\beta = 0.5$. Effect of the increased ratio $p/\nu$ on the velocity field in two regions in the fluid depth (the distance measured from the plate ( $y = 0$ ). Example 1.

  a) Near field – 2D profiles $u/U_0 = f(p/\nu)$
  b) Far field– 2D profiles $u/U_0 = f(p/\nu)$
  c) Short-time velocity field ($t = 0.1$ : 3D representation $u/U_0 = f(p/\nu, y)$

Medium times velocity field ($t = 1.0$ : 3D representation $u/U_0 = f(p/\nu, y)$

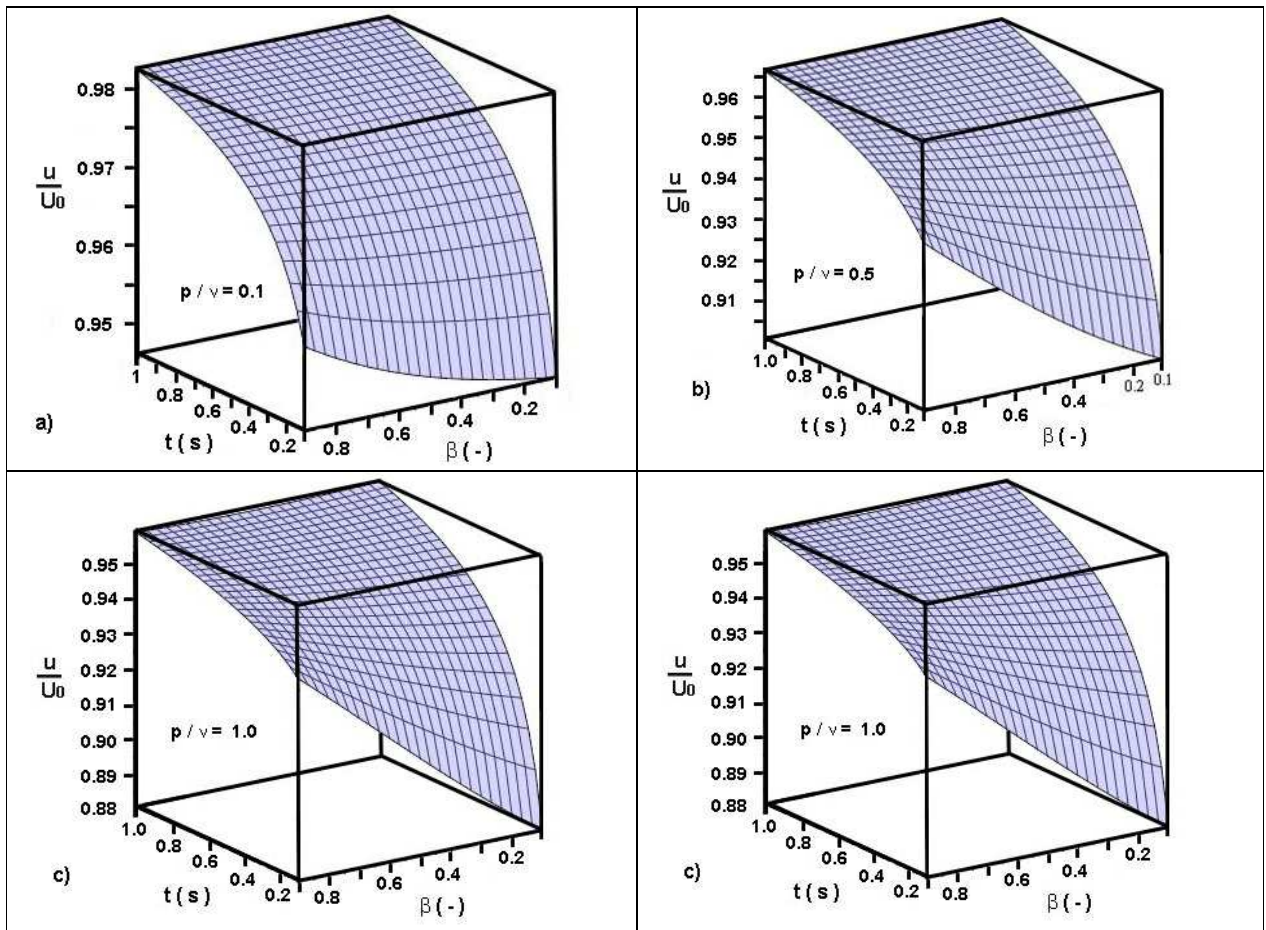

Fig.5. Velocity field as a function of the time ( $0.1 \leq t \leq 1.0$ ) and the and the fractional order $\beta$ for various ratios $p/\nu$ and $y = 0.5$. The penetration depth expressed by eq. (22) defines the approximate profile with $n = 2$. Example 1.

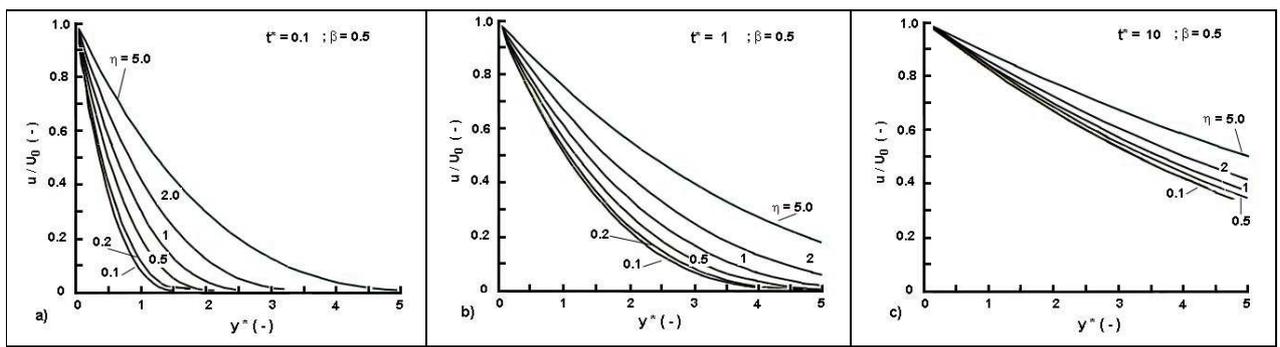

Fig.6. Dimensionless velocity profile (Example 2) expressed by the profile employing the penetration depth expressed by Eq.(25c) for three different dimensionless times $t^* = tU_0^2/\nu$ and various values of the parameter $\eta = \dfrac{(p/\nu)}{(\nu/U_0^2)} = (p/\nu)t^* = \dfrac{\alpha_1 U_0^2 \rho}{\mu^2}$ ..